# Automated Assignment of Backbone Resonances Using Residual Dipolar Couplings Acquired from a Protein with Known Structure

P. Shealy[1], R. Mukhopadhyay[1], S. Smith[1], and H. Valafar[1]

[1]Department of Computer Science and Engineering, University of South Carolina, Columbia, SC

**Abstract** – *Resonance assignment is a critical first step in the investigation of protein structures using NMR spectroscopy. The development of assignment methods that require less experimental data is possible with prior knowledge of the macromolecular structure. Automated methods of performing the task of resonance assignment can significantly reduce the financial cost and time requirement for protein structure determination. Such methods can also be beneficial in validating a protein's solution state structure. Here we present a new approach to the assignment problem. Our approach uses only RDC data to assign backbone resonances. It provides simultaneous order tensor estimation and assignment. Our approach compares independent order tensor estimates to determine when the correct order tensor has been found. We demonstrate the algorithm's viability using simulated data from the protein domain 1A1Z.*

## 1 Introduction

Structural genomics has produced a wealth of high-quality protein structures from NMR spectroscopy and X-ray crystallography. For proteins of known structure, NMR spectroscopy provides powerful techniques for further investigations of their biomedical and biophysical properties. For example, it becomes possible to study protein interactions with RNA, DNA, or other proteins, as well as protein kinematics. However, these studies require partial or complete resonance assignments.

A separate problem, determination of a protein's backbone structure, can also be performed more quickly if a close structural homologue is known. Structure determination is an important step in understanding the molecular basis for diseases and intelligent drug design. A significant portion of the data collected for structure determination by NMR is used for assignment. Once assignment is in place, a very small portion is used for elucidation of backbone structure. Elimination of the assignment step, given a known homologous structure, can therefore reduce the cost of protein structure production.

Here we introduce a new assignment method using only RDCs, with the goal of increasing assignment accuracy and reducing data collection times. We show that when using RDCs from multiple alignment media, we obtain highly accurate results. Furthermore, we rely solely on RDCs, simplifying the data collection step for resonance assignment.

## 2 Data and Method

### 2.1 Residual Dipolar Coupling

Residual dipolar couplings (RDCs) are values observed when a partial alignment is introduced in a molecule in the presence of an external magnetic field. RDCs are normally reduced to zero in an aqueous solution due to a molecule's isotropic tumbling. Introducing partial alignment results in non-zero RDCs. Partial alignment can be the result of an alignment medium [1], a binding tag [2], or a molecule's own magnetic susceptibility [3]. The RDC for an internuclear vector is a function of the angle it forms with the external magnetic field. Eq. (1) gives the time averaged RDC value for a vector between spin ½ nuclei.

$$D_{ij} = \frac{-\mu_0 \gamma_i \gamma_j h}{(2\pi r)^3} \left\langle \frac{3\cos^2(\theta_{ij})-1}{2} \right\rangle \quad (1)$$

Here, $D_{ij}$ is the residual dipolar coupling between the interacting nuclei $i$ and $j$ in units of Hz, $\gamma_i$ and $\gamma_j$ are their corresponding nuclear gyromagnetic ratios, $r$ is the internuclear vector length, and $\theta_{ij}$ is the angle between the internuclear vector and the external magnetic field. The angle brackets denote time-averaging.

RDCs are often collected for directly bonded atoms, such as backbone N-H, C'-N, and C'-C$^\alpha$. Recently, RDCs have been analyzed to provide information about protein structure [3-5], analyze molecular dynamics [5; 6], and assist in studies of carbohydrates [7; 8], nucleic acids [9; 10], and proteins [8; 11-16].

The *resonance assignment* problem, addressed in this paper, is that of completely assigning resonances to a protein of known structure. The 3D structures for many molecules are already known, often from X-ray crystallography, yet NMR assignments for these proteins are unavailable. Equipped with either partial or complete NMR assignment information, researchers can further investigate protein dynamics as well as protein interactions with DNA, RNA, other proteins, and small ligands.

| Medium | Order Parameters | | | | | | | |
|---|---|---|---|---|---|---|---|---|
| | α | β | γ | $S_{xx}$ | $S_{yy}$ | $S_{zz}$ | $D_a$ | R |
| 1 | 0 | 0 | 0 | 3e-4 | 5e-4 | -8e-4 | -4e-4 | 0.17 |
| 2 | 40 | 50 | 60 | -4e-4 | -6e-4 | 10e-4 | 5e-4 | 0.13 |
| 3 | -50 | 280 | 50 | 6e-4 | 9e-4 | -15e-4 | -7.5e-4 | 0.13 |

Figure 1: Order tensor parameters used to generate synthetic data for three independent alignment media. α, ß, and γ are Euler rotations about the Z, Y, and Z axes.

Previous assignment work has focused on resonance assignment using RDCs collected from multiple internuclear vectors [17], RDCs from multiple vectors in conjunction with amino acid specific chemical shifts [18], or RDCs in conjunction with amide exchange rates and NOEs [19]. Each method has produced results with varying accuracy.

## 2.2 Algorithm

We present an algorithm for assigning resonances using RDCs from backbone N-H vectors. In addition, using only N-H vectors avoids expensive $^{13}$C enrichment. Because our algorithm employs an exhaustive permutation technique (O(n!) time complexity), it is important to separate the data acquired from a protein into smaller subsets of data. We take advantage of the availability of a large set of techniques for isolating resonances by residue type in order to address this problem.

Our algorithm:

1. Collect residual dipolar couplings from a single alignment medium.
2. Isolate the RDCs belonging to each residue type.
3. Assign the vectors (and RDCs) for each residue type to a separate vector pool. Combine pools for residue types with few vectors until all pools have a sufficient number of vectors for step 4.
4. Permute the RDCs for each vector pool among all vectors in the pool and compute the error between the candidate RDC assignment and back-calculated RDCs. Retain the permutation with the lowest error as the correct permutation.
5. If the order tensor estimates from the pools are similar, terminate.
6. Otherwise, collect RDCs from another alignment medium, isolate the RDCs for each residue type, add the RDCs to the correct pool, and go to step 4.

Our algorithm uses RDCs collected in multiple alignment media and their categorization based on residue type. The identification of amino acid type of residues can be accomplished in a variety of ways, such as analysis of $C^\alpha$ and $C^\beta$ chemical shifts [20; 21], amino acid-specific labelings [22; 23], or a variety of NMR experiments [24; 25]. After combining pools with sparse vectors, the RDCs for each of the resulting pools are permuted to examine all possible RDC assignments. For each candidate assignment, the order tensor is estimated using the Singular Value Decomposition (SVD) [26], and the RMS error between the back-computed RDCs (using the known structure and order tensor) and unassigned RDCs is evaluated, as defined in Eq. (2). In this equation, $i$ denotes the residue number, $N$ denotes the number of residues for the pool, $m$ and $M$ denote the alignment medium number and number of media, respectively, and $c_i^m$ and $e_i^m$ are the $i^{th}$ experimental and back-computed RDCs for alignment medium $m$. This equation retains the pairing of RDCs across alignment media and is a Euclidean distance. The permutation with the lowest error is retained as the correct candidate assignment.

$$RMS = \frac{1}{N}\sum_{i=1}^{N}\sqrt{\sum_{m=1}^{M}(c_i^m - e_i^m)^2} \qquad (2)$$

Order tensor estimation is performed by solving the system of linear equations in Eq. (3) using the singular value decomposition. In this equation, $x_i$, $y_i$ and $z_i$ are the Cartesian coordinates of the $i^{th}$ internuclear vector, and $S_{xx}$, $S_{yy}$, $S_{xy}$, $S_{xz}$, and $S_{yz}$ are elements of the Saupe order tensor matrix [27].

$$\begin{bmatrix} x_1^2-z_1^2 & y_1^2-z_1^2 & 2x_1y_1 & 2x_1z_1 & 2y_1z_1 \\ \vdots & \vdots & \vdots & \vdots & \vdots \\ x_n^2-z_n^2 & y_n^2-z_n^2 & 2x_ny_n & 2x_nz_n & 2y_nz_n \end{bmatrix} \begin{bmatrix} S_{xx} \\ S_{yy} \\ S_{xy} \\ S_{xz} \\ S_{yz} \end{bmatrix} = \begin{bmatrix} r_1 \\ \vdots \\ r_n \end{bmatrix} \qquad (3)$$

A minimum of five RDCs from independent

|  | Alignment Media | | | | | | |
|---|---|---|---|---|---|---|---|
| *Vector Pool* | *1* | *2* | *3* | *1 & 2* | *1 & 3* | *2 & 3* | *1, 2, & 3* |
| *1* | 12 | 2321 | 27 | 0 | 1 | 0 | 0 |
| *2* | 8 | 0 | 0 | 0 | 0 | 0 | 0 |
| *3* | 52 | 6 | 0 | 0 | 0 | 0 | 0 |
| *4* | 23 | 194 | 0 | 0 | 0 | 0 | 0 |
| *5* | 2 | 8 | 0 | 0 | 0 | 0 | 0 |
| *6* | 0 | 4 | 0 | 0 | 0 | 0 | 0 |
| *7* | 0 | 0 | 0 | 0 | 0 | 0 | 0 |

Figure 2: The number of delusory permutations in each combination of alignment media, by vector pool.

vectors are required for order tensor estimation with Eq. (3). In the presence of noise, additional RDCs are incorporated to provide robustness. This is the main impetus in combining the RDCs for residue types with few vectors in order to obtain sets of a sufficient size for order tensor estimation. Pools of eight or more RDCs were sufficient for our data, which contains errors of ±1Hz. Vectors are pooled to maximize separation between RDCs, providing more robust order tensor estimation and assignment.

The algorithm compares order tensor estimates across vector pools to determine when the correct order tensor and assignment has been located for all vectors. Any fragment of a rigid molecule will exhibit the same order tensor. When there is insufficient data to reliably estimate the order tensor and assignments, the order tensors for different vector pools should vary widely. When sufficient data is present, order tensor estimates across vector pools will converge to the true order tensor. Here we visually compare order tensors using Sauson-Flamsteed plots, one per medium, containing order tensor estimates for all vector pools.

## 3 Results

We applied our method to the 83-residue death-effector domain with PDB code 1A1Z. This domain is primarily alpha-helical, with 60 of its residues contained in one of its five alpha helices. We computed simulated data using REDCAT [28], using the order tensor parameters given in Figure 1 to simulate three independent alignment media. We used REDCAT to extract N-H internuclear vectors and compute RDCs in each alignment medium. We added uniform noise in the range ±1Hz to the RDCs to simulate various sources of error, such as experimental error.

We distinguish between two types of assignment errors. A pseudoassignment error arises when an RDC is incorrectly assigned to an internuclear vector but lies within the error bounds of the vector's true RDC (e.g., assigning 10.2Hz to a vector with an RDC of 10.1Hz and 1Hz error). A *pseudoerrror permutation* is a permutation containing only pseudoassignments that has an error lower than the correct permutation's error. These permutations are acceptable when performing assignment. A *delusory permutation* is a permutation containing at least one true assignment error (not a pseudoassignment) that has an error lower than the correct permutation's error. We show in section 3.1 that delusory permutations are not uncommon for data collected under realistic experimental conditions.

### 3.1 Single Alignment Medium

Data from a single alignment medium proved insufficient to reliably distinguish the correct permutation. However, this failure is precedented by observing the degenerate nature of the RDC interaction [3; 29]. All three media, when examined individually, exhibited delusory permutations (Figure 2) and order tensor mis-estimates (Figure 3).

Media 1 and 2 had a substantial number of delusory permutations for five of the seven vector pools. In addition, the majority of the order tensor estimates from each medium differed substantially from the actual values. The estimates varied widely across the vector pools, suggesting that this problem is easily identified when the correct assignments are unknown.

Medium 3 exhibited the best performance. It contained delusory permutations only for vector pool 1. Most order tensor estimates were reasonably accurate, with the exception of pool 1, which varied significantly from the others.

### 3.2 Multiple Alignment Media

Our strategy in increasing the number of RDCs per residue is to collect RDCs in multiple alignment media. The unassigned RDCs can then be associated with a single (unknown) residue using chemical shifts. When permuting RDCs among residues, the RDCs always remain associated.

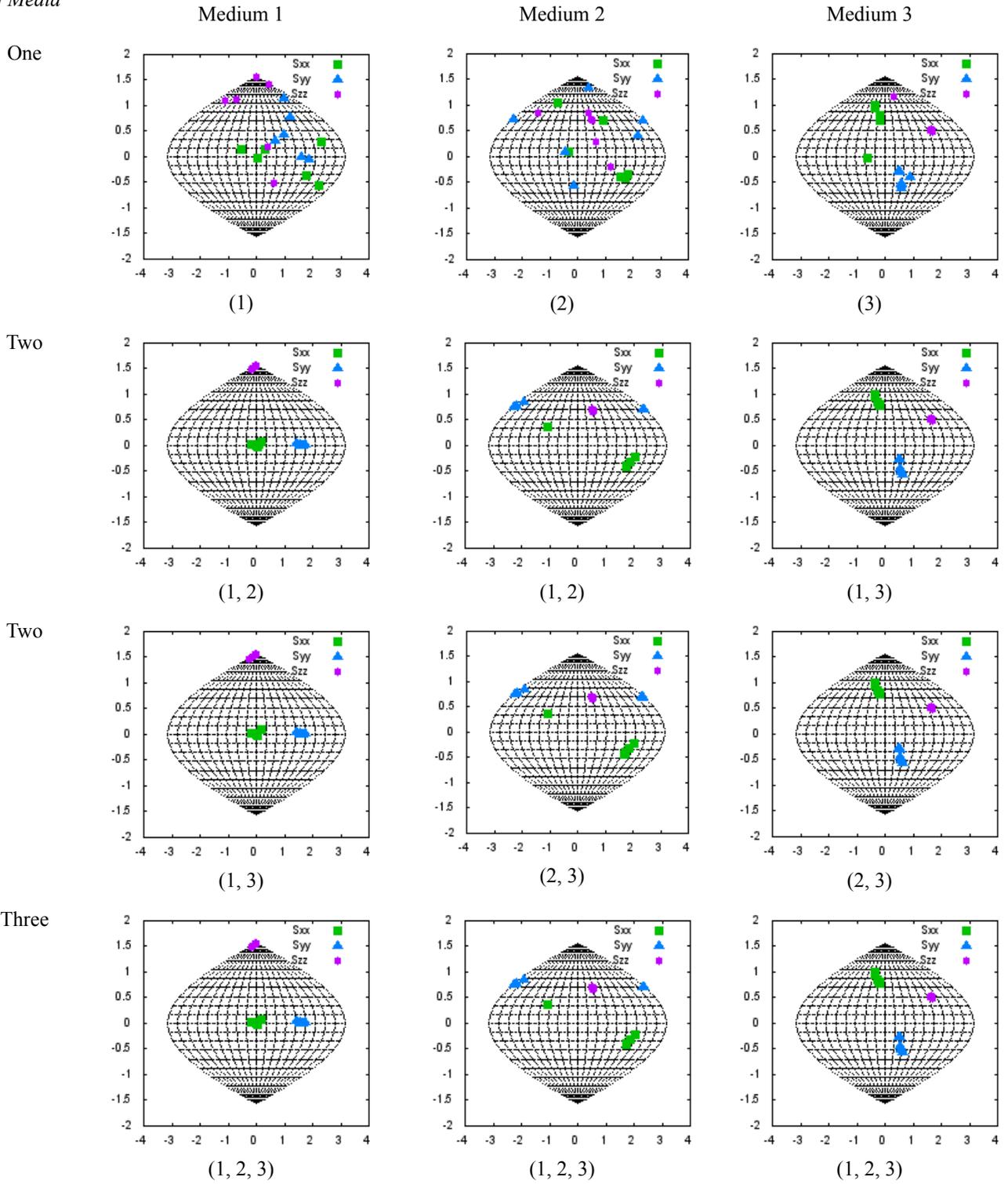

Figure 3: Sauson-Flamsteed plots of the orientational component of the order tensor estimates using RDCs from one, two, and three alignment media. The values in parentheses indicate the alignment media that were combined to generate the estimate.

| Order Tensor Parameter | Estimated | True |
|---|---|---|
| $S_{xx}$ | -2.33e-5 | -2.97e-5 |
| $S_{yy}$ | -5.74e-5 | -6.07e-5 |
| $S_{zz}$ | 8.07e-5 | 9.04e-5 |
| $S_{xy}$ | 4.130e-4 | 4.083e-4 |
| $S_{xz}$ | 6.472e-4 | 6.273e-4 |
| $S_{yz}$ | 4.533e-4 | 4.398e-4 |

Figure 4: Example estimated order tensor parameters for medium 3, using RDCs from media 1, 2, and 3.

Using two alignment media proved sufficient to obtain reasonable, if not perfect, order tensor estimates (Figure 3) and assignments (Figure 2). Among the three possible combinations of two media, a total of five permutations had errors lower than the correct permutation (one in alignment media 1 & 2, one in media 2 & 3, two in media 1 & 3). Four of these permutations were pseudoassignments, leaving only a single delusory permutation in pool 1. Furthermore, this delusory permutation contains only a single error that is not a pseudoassignment error. This error is an estimated RDC of -0.57 assigned to a vector with an experimental RDC of -0.93 and true RDC of -1.92 (in medium 1).

Three alignment media were sufficient to eliminate all delusory permutations. In addition, no pseudoerror permutations appeared. Furthermore, using RDC data from three alignment media also increased the separation between the RMS errors of the first (i.e., correct) and second permutations.

The vector pools were formed to maximize separation between RDCs, while ensuring that the resulting pools were not too large to permute in a reasonable amount of time. The maximum plausible pool size in our experiments was 12 vectors. We compare order tensor estimates by examination of Sauson-Flamsteed plots with all vector pools, taking into account rotational degeneracies.

## 4 Discussion

We have introduced an algorithm for resonance assignment. It is designed to reduce data collection times by utilizing only residual dipolar couplings from multiple alignment media. The formulation presented allows simultaneous order tensor estimation and assignment.

Assignment with alpha-helical proteins such as 1A1Z is difficult because each alpha helix contains N-H vectors that are nearly parallel. These vectors have similar RDCs, so distinguishing between the correct assignments is difficult. Given our success in assigning resonances for 1A1Z, we expect our method to be widely applicable.

Our algorithm leverages the availability of a broad set of techniques for isolating resonances by vector type. This reduces the set of possible resonance assignments substantially. Separating resonances into several vector pools has the additional advantage of providing a reliable estimate for when the correct order tensor has been identified.

Under experimental conditions, when the correct assignments are unknown, a permutation algorithm with a single alignment medium will likely produce incorrect assignments. 1A1Z exhibited substantial difficulty with media 1 and 2, and medium 3 failed to produce satisfactory assignments for all vector pools. Using multiple vectors in a single alignment medium also proved insufficient to fully assign resonances using RDCs, which is consistent with previous results.

Data from two independent alignment media were sufficient for 1A1Z. Even though the combined data from media 1 & 2 contained a delusory permutation, the difference between the estimated and true RDC is small compared to the range of observable RDCs from N-H vectors in medium 1, and the error can be partially explained by an experimental RDC that is closer to to the estimate than the true value.

Using three alignment media proved sufficient to eliminate all assignment errors, including pseudo assignments. In addition, the order tensor estimates were reliably estimated for all three alignment media by averaging across vector pools. An example is shown in Figure 4, where the order tensor parameters for medium 2 were estimated by averaging each parameter across the vector pools, using the best order tensor estimate from each pool. RDCs from multiple alignment media are useful in reducing the ambiguity about an internuclear vector's exact position [30]. In this light, it is not surprising that multiple alignment media perform better than multiple vectors in a single alignment medium.

Order tensor estimates are compared across vector pools to determine when the correct assignment has been found. Under realistic experimental conditions, when the correct assignments are unknown, this independent measure provides a reliable estimate of when the final answer has been located. As Figure 3 demonstrates, estimating order tensor parameters with an insufficient number of alignment media results in widely varying estimates across vector pools.

## 4.1 Conclusion and Future Directions

One of the limiting factors in applying the algorithm to larger proteins is the permutation step. For 1A1Z, data for leucines was omitted because 1A1Z contains 21 residues of this type, which is too many to permute in an acceptable time period. Although the order tensor estimate from other pools can guide assignment in this case, we aim to investigate algorithmic alternatives to considering all permutations.

We hope to apply our algorithm to experimental data. Doing so will require extending the algorithm to handle experimental conditions, such as missing resonances. We also hope to incorporate additional types of data when it is available, such as chemical shift connectivity information.

Finally, we hope to integrate PDPA [30] into the assignment algorithm. PDPA provides an estimate of the structure's order tensor parameters and assignment of RDCs to internuclear vectors from unknown residues. These estimates can then be used to initialize a modified version of the assignment algorithm.

## 5 References

[1] Prestegard JH, Valafar H, Glushka J & Tian F. Nuclear magnetic resonance in the era of structural genomics. *Biochemistry* (2001) 40: pp. 8677-8685.
[2] Nitz M, Sherawat M, Franz KJ, Peisach E, Allen KN & Imperiali B. Structural origin of the high affinity of a chemically evolved lanthanide-binding peptide. *Angew Chem Int Ed Engl* (2004) 43: pp. 3682-3685.
[3] Prestegard JH, al-Hashimi HM & Tolman JR. NMR structures of biomolecules using field oriented media and residual dipolar couplings. *Q Rev Biophys* (2000) 33: pp. 371-424.
[4] Bax A, Kontaxis G & Tjandra N. Dipolar couplings in macromolecular structure determination. *Methods Enzymol* (2001) 339: pp. 127-174.
[5] Bryson M, Tian F, Prestegard JH & Valafar H. REDCRAFT: A tool for simultaneous characterization of protein backbone structure and motion from RDC data. *J Magn Reson* (2008) : .
[6] Tolman JR, Al-Hashimi HM, Kay LE & Prestegard JH. Structural and dynamic analysis of residual dipolar coupling data for proteins. *J Am Chem Soc* (2001) 123: pp. 1416-1424.
[7] Azurmendi HF, Martin-Pastor M & Bush CA. Conformational studies of Lewis X and Lewis A trisaccharides using NMR residual dipolar couplings. *Biopolymers* (2002) 63: pp. 89-98.
[8] Tian F, Valafar H & Prestegard JH. A dipolar coupling based strategy for simultaneous resonance assignment and structure determination of protein backbones. *J Am Chem Soc* (2001) 123: pp. 11791-11796.
[9] Tjandra N, Tate S, Ono A, Kainosho M & Bax A. The NMR Structure of a DNA Dodecamer in an Aqueous Dilute Liquid Crystalline Phase. *Journal of the American Chemical Society* (2000) 122: pp. 6190-6200.
[10] Vermeulen A, Zhou H & Pardi A. Determining DNA Global Structure and DNA Bending by Application of NMR Residual Dipolar Couplings. *Journal of the American Chemical Society* (2000) 122: pp. 9638-9647.
[11] Assfalg M, Bertini I, Turano P, Grant Mauk A, Winkler JR & Gray HB. 15N-1H Residual dipolar coupling analysis of native and alkaline-K79A Saccharomyces cerevisiae cytochrome c. *Biophys J* (2003) 84: pp. 3917-3923.
[12] Bertini I, Luchinat C, Turano P, Battaini G & Casella L. The magnetic properties of myoglobin as studied by NMR spectroscopy. *Chemistry* (2003) 9: pp. 2316-2322.
[13] Clore GM & Bewley CA. Using conjoined rigid body/torsion angle simulated annealing to determine the relative orientation of covalently linked protein domains from dipolar couplings. *J Magn Reson* (2002) 154: pp. 329-335.
[14] Cornilescu G, Delaglio F & Bax A. Protein backbone angle restraints from searching a database for chemical shift and sequence homology. *J Biomol NMR* (1999) 13: pp. 289-302.
[15] Fowler CA, Tian F, Al-Hashimi HM & Prestegard JH. Rapid determination of protein folds using residual dipolar couplings. *J Mol Biol* (2000) 304: pp. 447-460.
[16] Wang L & Donald BR. Exact solutions for internuclear vectors and backbone dihedral angles from NH residual dipolar couplings in two media, and their application in a systematic search algorithm for determining protein backbone structure. *J Biomol NMR* (2004) 29: pp. 223-242.
[17] Zweckstetter M. Determination of molecular alignment tensors without backbone resonance assignment: Aid to rapid analysis of protein-protein interactions. *J Biomol NMR* (2003) 27: pp. 41-56.
[18] Hus J, Prompers JJ & Bruschweiler R. Assignment Strategy for Proteins with Known Structure. *Journal of Magnetic Resonance* (2002) 157: pp. 119-123.
[19] Langmead CJ, Yan A, Lilien R, Wang L & Donald BR. A polynomial-time nuclear vector replacement algorithm for automated NMR resonance assignments. *J Comput Biol* (2004) 11: pp. 277-298.
[20] Grzesiek S & Bax A. Amino acid type determination in the sequential assignment procedure of uniformly 13C/15N-enriched proteins. *J Biomol NMR* (1993) 3: pp. 185-204.
[21] Spera S & Bax A. Empirical correlation between protein backbone conformation and C.alpha. and C.beta. 13C nuclear magnetic resonance chemical shifts. *Journal of the American Chemical Society* (1991) 113: pp. 5490-5492.
[22] Gronenborn AM & Clore GM. Rapid screening for structural integrity of expressed proteins by heteronuclear NMR spectroscopy. *Protein Sci* (1996) 5: pp. 174-177.
[23] Ou HD, Lai HC, Serber Z & Dotsch V. Efficient

identification of amino acid types for fast protein backbone assignments. *J Biomol NMR* (2001) 21: pp. 269-273.
[24] Dotsch V, Oswald RE & Wagner G. Amino-acid-type-selective triple-resonance experiments. *J Magn Reson B* (1996) 110: pp. 107-111.
[25] Schubert M, Oschkinat H & Schmieder P. MUSIC and aromatic residues: amino acid type-selective (1)H-(15)N correlations, III. *J Magn Reson* (2001) 153: pp. 186-192.
[26] Losonczi JA, Andrec M, Fischer MWF & Prestegard JH. Order Matrix Analysis of Residual Dipolar Couplings Using Singular Value Decomposition. *Journal of Magnetic Resonance* (1999) 138: pp. 334-342.
[27] Saupe, A. & Englert, G.. High-Resolution Nuclear Magnetic Resonance Spectra of Orientated Molecules. *Phys. Rev. Lett.* (1963) 11: p. pp. 462-464.
[28] Valafar H & Prestegard JH. REDCAT: a residual dipolar coupling analysis tool. *J Magn Reson* (2004) 167: pp. 228-241.
[29] Al-Hashimi HM, Valafar H, Terrell M, Zartler ER, Eidsness MK & Prestegard JH. Variation of molecular alignment as a means of resolving orientational ambiguities in protein structures from dipolar couplings. *J Magn Reson* (2000) 143: pp. 402-406.
[30] Miao X. Homologue Protein Structure Detection using Unassigned Residual Dipolar Coupling Data from Multiple Alignment Media. University of South Carolina. 2007.